\documentclass[%
preprint,
aps,
pre,
amsmath,amssymb,
]{revtex4-2}

\usepackage{graphicx}%
\usepackage{dcolumn}%
\usepackage{wrapfig}
\usepackage{multirow}
\usepackage{subfig}
\graphicspath{{Figures/}}
\usepackage{float}
\usepackage{dcolumn}%
\usepackage{bm}%
\usepackage[hidelinks]{hyperref}%
 \hypersetup{
     colorlinks=true,
     linkcolor=blue,
     filecolor=blue,
     citecolor = blue,      
     urlcolor=cyan,
     }

\begin{document}

\title{Embedding theory contributions to average atom models for warm dense matter}

\author{Sameen Yunus}
\author{David A. Strubbe}
 \email{dstrubbe@ucmerced.edu}
\affiliation{%
 Department of Physics, University of California, Merced, 5200 N. Lake Rd, Merced
 }%

\date{\today}

\begin{abstract}

Accurate modeling in the warm dense matter regime is a persistent challenge with the most detailed models such as quantum molecular dynamics and path integral Monte Carlo being immensely computationally expensive. Density functional theory (DFT)-based average atom models (AAM) offer significant speed-ups in calculation times while still retaining fair accuracy in evaluating equations of state, mean ionizations, and more. Despite their success, AAMs struggle to precisely account for electronic interactions -- in particular, they do not account for effects on the kinetic energy arising from overlaps in neighboring atom densities. We aim to enhance these models by including such interactions via the non-additive kinetic potential $v^{\rm nadd}$ as in DFT embedding theories. $v^{\rm nadd}$ can be computed using Thomas-Fermi, von Weizs\"acker, or more sophisticated kinetic energy functionals. The proposed model introduces $v^{\rm nadd}$ as a novel interaction term in existing ion-correlation models, which include interactions beyond the central atom. We have applied this model to hydrogen at 5 eV and densities ranging 0.008 to 0.8 g/cm$^3$, and investigated the effects of $v^{\rm nadd}$ on electron densities, Kohn-Sham energy level shifts, mean ionization, and total energies. 

\end{abstract}

\maketitle

\section{\label{sec:Intro}Introduction}

Accurate models of warm dense matter (WDM) can lead to an improved understanding of material properties in conditions relevant to many important phenomena. Perhaps most pressing of which is the modeling of inertial confinement fusion as fuel is compressed and heated on its way to ignition \cite{Kritcher2022, NIF_ignition2024, Atzeni2004}. Warm dense plasmas also occur in the cores of giant planets \cite{Nettelmann2008}, brown dwarfs \cite{Booth2015}, and white dwarf envelopes \cite{Saumon2022} leading to novel astrophysics that can be studied in laboratories and through simulations. Studying warm dense matter is challenging due to the strong degree of Coulomb coupling and quantum degeneracy that needs to be accounted for in simulations. Naturally, the complex intertwining of such physics poses a theoretical challenge in what kinds of models can accurately describe WDM -- here density functional theory-based average atom models \cite{Liberman1979, Starrett2013} have been very successful at providing accurate and computationally efficient models of the complex dense plasma environment. We briefly introduce the current state of average atom models and a contribution in the electronic structure that can improve existing models.

Average atom models (AAM) alleviate the computational expense of detailed \textit{ab initio} molecular dynamics models \cite{Zerah1992,Lambert2011,Ruter2014,Krypton_QMD2020} by considering spherically averaged ion spheres each of which contain a central nucleus and an associated electron density thus reducing the full electronic structure calculation to that for a single atom with spherical symmetry. The plasma environment is partitioned according to the mean ion density, $n_\text{I}^0$, by the Wigner-Seitz radius and the resulting average atom can be solved to produce accurate electronic properties:
\begin{equation}
R_\text{WS} = \left( \frac{3}{4\pi n_\text{I}^0}\right)^{1/3}.
\end{equation}

The earliest AAMs \cite{WS1934,TFD1949,TFD1955} were built on the Thomas-Fermi approximation \cite{Thomas1927,Fermi1927} of continuous electron density for treating extreme conditions and metallic electrons. These provided analytic solutions for such electron densities but had no description of states -- these were introduced in the first Kohn-Sham (KS) DFT-based AAM by Liberman in 1979 \cite{Liberman1979}. This model has been very successful in describing dense plasma electronic properties and can reproduce shell structure in the density as opposed to the earlier density-only Thomas-Fermi-like prescriptions. The interacting ions in the plasma environment are approximated as spherically symmetric ion spheres of size $R_\text{WS}$ and the spherical symmetry is leveraged to solve single-coordinate radial KS equations. This class of ``ion sphere models" are applied widely for evaluating equations of state in the WDM regime \cite{Liberman1979,Renaudin2003,Clerouin2012,Faussurier2019}. 

It should be noted that while these are very efficient and fairly accurate models, they ignore inter-cell interactions between ion spheres and these are only partially accounted for through boundary conditions, which distinguish the ion sphere model from purely isolated atom models. These include interactions with neighboring atoms that include some limited details of electron-electron interactions between neighboring cells but do not in general contain any information about ion-correlations. The ion-sphere model is fairly robust at relatively high temperatures and low densities, but it becomes less accurate as density increases and electron-electron and electron-ion interaction effects become more important. In addition, ion sphere models are also subject to arbitrary choices of boundary conditions at the Wigner-Seitz radius which impacts observables in the calculation \cite{CallowPRR2022}.

Several models have been introduced to include correlations beyond the central atom, known as ``ion correlation models." The strength of these models lies in their ability to incorporate correlations beyond the ion sphere, with pair correlation functions evaluated either through a semi-classical hyper-netted chain model \cite{DW_Perrot1982, Rozsnyai1991} or self-consistently through plasma closure relations \cite{Starrett2012}. These treat a larger correlation sphere of size $R_\text{corr}$, typically 5-6 times the Wigner-Seitz radius \cite{Murillo2013}. This radius is chosen such that the ion-ion correlations ($g_\text{II}$) and ion-electron correlations ($g_\text{Ie}$), which describe the distribution of particles around a reference particle, approach unity around $r \lesssim R_\text{corr}$ \cite{DW_Perrot1982} to include the long-range interactions.

We propose a re-framing of the ion-correlation models as an embedding theory problem where a central reference atom is embedded in an averaged background plasma contribution based on ion-ion pair-correlation functions $g_\text{II}$. This introduces a new contribution in the electronic structure of the average atom which is the non-additive kinetic potential $v^\text{nadd}$ arising from the orthogonality condition of overlapping orbitals between the embedded system and the environment. We use the Mermin finite-temperature DFT formalism \cite{Hohenberg-Kohn1964, Kohn-Sham1965, Mermin1965} to construct such an embedding theory-average atom model and study the contributions to electron densities and KS eigenvalues in the Octopus real-space DFT code \cite{Octopus2015,Octopus2020}. The main goals of this paper are to explore how this embedding contribution changes DFT observables compared to an average atom model without $v^\text{nadd}$. The model is presented in Sec. \ref{sec:Model}, with a brief introduction to embedding theory and approaches to calculating $v^\text{nadd}$ in Secs. \ref{sec:Vnad} and \ref{sec:KEDFs}. The practical implementation to warm dense hydrogen is discussed in Sec. \ref{sec:practical} with treatment of continuum electrons briefly described in Sec. \ref{sec:Continuum} and other computational parameters in Sec. \ref{sec:parameters}. Finally, in Sec. \ref{sec:Results} we apply the model to present some effects arising due to the embedding contribution, and then conclude in Sec. \ref{sec:Conclusion}.

\section{\label{sec:Model}The proposed model and parameters}

\subsection{\label{sec:Vnad} Embedding theory and the non-additive kinetic potential}

Embedding theory is a general method used in chemistry to derive detailed properties of molecules embedded in averaged solvent environments. This approach involves making a detailed quantum mechanical treatment of the central molecule or protein while considering an average potential of the surrounding solvent. In particular, we are considering frozen density embedding theory \cite{Wesolowski1993, Wesolowski2015, Polak2022} in which the full density is partitioned into the embedded subsystem and the environment. This is somewhat reminiscent of the ion correlation models \cite{DW_Perrot1982} where the central ion sphere is embedded in a large correlation sphere which acts as the average environment potential of the plasma. However, these models do not typically consider the electrons as an embedded subsystem and neglect a non-additive contribution to their kinetic energy that comes from partitioning the density between the central atom and environment -- this is the non-additive kinetic potential bi-functional or $v^\text{nadd}$ \cite{Wesolowski1993}. Starting from the ordinary KS equations for the electron density of single particles in the full system,
\begin{equation}
\left[ -\frac{\nabla^2}{2} + V_{\text{ext}}(\bm{r}) + V_{\text{H}}[n^{\text{tot}}](\bm{r}) + 
V_{\text{xc}}[n^\text{tot}(\bm{r})]
 \right] \phi_i(\bm{r}) = \epsilon_i \phi_i(\bm{r}),
\end{equation}
we partition the total density of the system into the embedded subsystem and the environment as $n_\text{tot}=n_\text{sub}+n_\text{env}$:
\begin{eqnarray}
    \left[ -\frac{\nabla^2}{2} + V_\text{ext}^\text{sub}(\mathbf{r}) + V_\text{ext}^\text{env}(\mathbf{r}) + V_\text{H}[n^\text{sub} + n^\text{env}](\bm{r}) + 
    V_{\text{xc}}[n_\text{sub}+n_\text{env}](\bm{r}) + \right. \nonumber\\
    \left. v^{\text{nadd}}[ n^\text{sub}, n^\text{env}](\bm{r}) \right] \phi_i(\bm{r}) = \epsilon_i \phi_i(\bm{r})
\label{eq:embedding}    
\end{eqnarray}
The external potential ($V_{\rm ext}$), independent of density, and the Hartree ($V_{\rm H}$) and exchange-correlation ($V_{\rm xc}$) potentials, functionals of the density, are straightforward to partition. However, the partitioning of the exact kinetic energy of single-particle states into an environment and subsystem yields an embedding potential that must be approximated. In principle, if we had an exact expression for the kinetic-energy functional $T_s[n]$ we would obtain an exact orbital-free expression for the embedding potential, without any dependence on KS states. In Frozen Density Embedding Theory (FDET), the constrained search yields an exact orbital-free expression for such an embedding potential, relying solely on the electron density and no other properties of the environment \cite{Polak2022}:
\begin{equation}
v^{\text{nadd}}[ n^{\text{sub}}, n^{\text{env}}](\bm{r}) = 
\frac{\delta T^\text{nadd}_s[n^\text{sub}, n^\text{env}](\bm{r})}{\delta n^\text{sub}(\bm{r})}  =
\left.\frac{\delta T_s[n](\bm{r})}{\delta n(\bm{r})}\right|_{n=n^\text{sub}+n^\text{env}} - 
\left.\frac{\delta T_s[n](\bm{r})}{\delta n(\bm{r})}\right|_{n=n^\text{sub}}.
\label{eq:Vnad}
\end{equation}
This term arises from the functional derivative of the non-additive part of the kinetic energy bi-functional with respect to the subsystem density:
\begin{equation}
T^\text{nadd}_s[n^\text{sub}, n^\text{env}]=T^\text{nadd}_s[n^\text{sub}+ n^\text{env}]-T^\text{nadd}_s[n^\text{sub}] - T^\text{nadd}_s[n^\text{env}].
\end{equation}
In the limit of non-overlapping $n^\text{sub}$ and $n^\text{env}$, $T^\text{nadd}_s[n^\text{sub}, n^\text{env}]=0$ and we recover the exact kinetic energy of a non-interacting system \cite{Polak2022}.

An alternative, perhaps more intuitive way to conceptualize $v^\text{nadd}$, is to recognize that as we partition the system into overlapping densities, the orthogonality of eigenstates of the full system, due to the Pauli exclusion principle, must be maintained. This orthogonality constraint affects the curvature of the states especially near the regions of strong overlap which results in a contribution to the kinetic energy of the system that cannot be accounted for by the kinetic energies of states of the subsystems. Densities in the warm dense matter regime certainly enter into regions of strong orbital overlap, but average atom models do not have a prescription for explicitly including such effects in the electronic structure. Our goal with this work is to provide a $v^\text{nadd}$ correction to existing AAMs and explore how this affects DFT observables across a range of densities and temperatures. This approach provides an approximate way of including hybridization in AAMs, which is not taken into account in existing models.

\subsection{\label{sec:KEDFs} Kinetic energy functionals and their role in embedding theory}

In principle, $v^\text{nadd}$ should be exactly defined by an exact kinetic-energy functional $T_s [n]$; however, as with the XC functional, there is no known exact expression for this and approximations must be made (indicated by $\tilde{T}_s$). Kinetic energy functionals are much less developed than XC approximations but $\tilde{T}_s$ approximations are ubiquitous in the orbital-free DFT literature for WDM \cite{Zerah1992,Lambert2007,White2013,Karasiev2014}. The exact XC and kinetic energy functionals should include finite-temperature effects \cite{Karasiev2016,Perrot1979,Karasiev2012}, but our work so far uses only zero-temperature approximations to both. In our framework, we have applied the spherical symmetry of the average atom model to our embedding potential Eq. \ref{eq:Vnad}, which is reduced to just the radial coordinate $r$. We can now approximate the exact $v^\text{nadd}$ of Eq. \ref{eq:Vnad} with our choice of approximate kinetic-energy functionals, $\tilde{T}_s$:
\begin{equation}
\tilde{v}^{\text{nadd}}[ n^{\text{atom}}, n^{\text{env}}](r) = 
\left.\frac{\delta \tilde{T}_s[n](r)}{\delta n(r)}\right|_{n=n^\text{atom}+n^\text{env}} - 
\left.\frac{\delta \tilde{T}_s[n](r)}{\delta n(r)}\right|_{n=n^\text{atom}}.
\label{eq:Vnad_AAM}
\end{equation}
A common starting point is the Thomas-Fermi kinetic energy functional \cite{Thomas1927,Fermi1927} which is exact for the uniform electron gas and only depends on the local value of the density at any given point $\bm{r}$: 
\begin{equation}
T_s^{\text{TF}}[n(\bm{r})] = C^{\text{TF}} \int n(\bm{r})^{5/3} \, d\bm{r},  \qquad
C^{\text{TF}} = \frac{3}{10} (3\pi^2)^{2/3}
\end{equation}
The corresponding non-additive kinetic energy is then:
\begin{equation}
\tilde{T}_s^{\text{nadd(TF)}} [n^\text{atom}, n^\text{env}](r) =  C^\text{TF} \int (n^\text{atom}(r) + n^\text{env}(r))^{5/3} - (n^\text{atom}(r))^{5/3} - (n^\text{env}(r))^{5/3} \, d\bm{r},  
\end{equation}
The non-additive kinetic potential bi-functional $v^\text{nadd}$ is given by the functional derivative of $\tilde{T}_s^{\text{nadd(TF)}}$ with respect to the density of the embedded subsystem (the central atom in our case):
\begin{equation}
\begin{aligned}
    \tilde{v}^\text{nadd(TF)} [n^\text{atom}, n^\text{env}](r) = \frac{\delta}{\delta n^\text{atom}(r)} \tilde{T}_s^\text{nadd(TF)}[n^\text{atom}, n^\text{env}](r) \\
    = \frac{5}{3} C^\text{TF} \left[ (n^\text{atom}(r) + n^\text{env}(r))^{2/3} - (n^\text{atom}(r))^{2/3} \right]    
\end{aligned}
\end{equation}
Another common approximation with an analytical form is the von Weizs\"acker functional \cite{Weizsacker1935} which is exact for one-electron or two-electron spin-compensated systems. This functional is equivalent to the analytically inverted potential from a one-orbital KS equation \cite{Banafsheh2022} and allows for an easy test case for the numerical implementation of our functionals since they are analytically solvable for hydrogen in the ground state. The exact von Weizs\"acker kinetic functional
\begin{equation}
T_s^{\text{vW}}[n(\bm{r})] = \int \frac{|\nabla_{\bm{r}} n|^2}{8n} \, d\bm{r}
\end{equation}
yields the corresponding non-additive kinetic energy functional \cite{Wesolowski1993,Polak2022},
\begin{equation}
\begin{aligned}
    \tilde{T}_s^{\text{nadd(vW)}}[n^\text{atom}, n^\text{env}](r) = 
    \int \frac{|\nabla (n^\text{atom}+ n^\text{env})|^2}{8(n^\text{atom}+ n^\text{env})} \, dr - 
    \int \frac{|\nabla n^\text{atom}|^2}{8n^\text{atom}} \, dr - 
    \int \frac{|\nabla n^\text{env}|^2}{8n^\text{env}} \, dr \\
    =-\frac{1}{8} \int \frac{|n^\text{atom}\nabla n^\text{env} - n^\text{env}\nabla n^\text{atom}|^2}{n^\text{atom}n^\text{env}(n^\text{atom}+n^\text{env})} \, dr
\end{aligned}
\end{equation}
and non-additive kinetic potential functional as follows after some manipulation of gradients:
\begin{equation}
\begin{aligned}
    \tilde{v}^\text{nadd(vW)} [n^\text{atom}, n^\text{env}](r) = \frac{\delta}{\delta n^\text{atom}(r)} \tilde{T}_s^\text{nadd(vW)}[n^\text{atom}, n^\text{env}](r) \\
    = \frac{|\nabla (n^\text{atom}+n^\text{env})|^2}{8(n^\text{atom}+n^\text{env})^2} - \frac{\nabla^2 (n^\text{atom}+n^\text{env})}{4(n^\text{atom}+n^\text{env})} - 
    \frac{|\nabla n^\text{atom}|^2}{8(n^\text{atom})^2} + \frac{\nabla^2 n^\text{atom}}{4n^\text{atom}} 
\end{aligned}
\end{equation}
As mentioned, these are exact in the limits of a homogeneous electron gas (TF) and a one- or two-electron system (vW). The von Weizs\"acker functional is shown \cite{Dreizler1990} to provide a rigorous lower bound to the true kinetic energy and the Thomas-Fermi functional can be interpreted as a correction to this. For non-interacting particles in one spatial dimension ($r$) \cite{March1958} the sum of these provides a rigorous upper bound:
\begin{equation}
    T_s^\text{vW}[n] \leq T_s \leq T_s^\text{vW}[n] + T_s^\text{TF}[n].
\end{equation}
On the other hand, the density-gradient expansion \cite{Kirznits1956} yields the Thomas-Fermi term as the zeroth order with the von Weizs\"acker term as the second-order term of the gradient expansion with an additional $1/9$ coefficient:
\begin{equation}
\begin{aligned}
    T_s[n] = T_s^{(0)}[n] + T_s^{(2)}[n] + T_s^{(4)}[n] +... \\
    T_s[n] \simeq T_s^\text{TF}[n] + \frac{1}{9}T_s^\text{vW}[n]
\end{aligned}    
\end{equation}
This provides us a few functional approximations to explore for $\tilde{v}^\text{nadd}$ that are exact in some particular limits.

\subsection{\label{sec:practical}Implementation and application to hydrogen plasmas}

We applied this method to hydrogen at WDM conditions for its relevance to ICF fuels \cite{Lambert2011} and in planetary interiors \cite{Nettelmann2008,Helled2020}, and because hydrogen has analytical solutions at the ground state for many of the properties that we evaluated. In general, the $v^\text{nadd}$ contribution can be applied to an average atom model of any element with the use of pseudopotentials \cite{Phillips1959}. In fact, this model may prove to be more accurate for helium and beyond since the approximations used in DFT lead to some peculiarities for hydrogen that deviate from the true ground-state behavior for instance as discussed in Table 6.1 of Ref. \cite{DFTPrimer2003}. Extreme pressures in the WDM regime can result in core orbital overlaps between adjacent atoms, especially in the stagnation pressures present in laboratory fusion studies \cite{Kritcher2020, Kononov2023}. For this reason, it can be important to consider semi-core pseudopotentials or use harder pseudopotentials with small cutoff radii \cite{Recoules2009}. The region where deviations are important lies between the pseudopotential cutoff $r_\text{cut}$ and the Wigner-Seitz radius; radii smaller than $R_\text{WS}$ do not contribute to the environment since no other atoms are allowed in this region. For radii greater than $r_\text{cut}$, the hydrogen pseudopotentials are explicitly constructed such that the eigenenergies and pseudo-wavefunctions agree with the all-electron calculation. We compared the kinetic potential due to the atom $\left.\frac{\delta T_s}{\delta n}\right|_\text{atom}$ using the hydrogen pseudopotentials against an all-electron calculation using the Atomic Pseudopotential Engine \cite{APE2008} and found the kinetic potentials agree to within 10 meV; this implies that our use of pseudopotentials for the ion densities considered does not greatly impact the accuracy of our results. For larger densities, the results would benefit in accuracy from the use of smaller cutoff radius. We choose Octopus \cite{Octopus2015,Octopus2020} for our calculations: as a real-space code, it allows us to study finite spherical systems which is impractical with plane-wave codes, and it allows the use of a fully user-defined piece-wise external potential, which is how the constructed AAM is passed to Octopus. 

The model is introduced here by re-framing the embedding equation (Eq. \ref{eq:embedding}) in terms of the relevant quantities for the average atom model. We introduce the electron subscript in the electron density $n_\text{e}$ to distinguish it from the ion density $n_\text{I}$ which will become relevant shortly. Additionally, we can separate the terms in the total potential into the contributions due to the KS potential of the central atom; the external, Hartee, and exchange-correlation terms from the averaged plasma environment; and finally the non-additive $v^\text{nadd}$ term which accounts for the kinetic energy contribution arising from orbital overlaps between the atom and environment. It should be noted here that an unnecessary approximation has been made so far in linearizing the exchange-correlation contributions due to the embedded atom and environment, which will be corrected in further work. While this neglected non-additive XC term should be a small contribution overall, the approximation can lead to a slightly increased error in the DFT calculations of the total energy, densities, and KS states.

Our equation becomes:
\begin{eqnarray}
    &&\left[ -\frac{\nabla^2}{2} + V_{\text{KS}}^{\text{atom}}(\mathbf{r}) + 
    V_{\text{ext}}^{\text{env}}(\mathbf{r}) + V_\text{H}[n_{\text{e}}^{\text{env}}](\mathbf{r}) + V_\text{xc}[n^\text{env}] + \right. \nonumber \\
    &&\left. v^{\text{nadd}}[n_{\text{e}}^{\text{atom}}, n_{\text{e}}^{\text{env}}](\mathbf{r})
    \right] \phi_i(\bm{r}) = \epsilon_i \phi_i(\bm{r}), 
    \label{eq:AAM Vtot}
    \\
    &&V_{\text{KS}}^{\text{atom}}(\mathbf{r}) = V_{\text{ext}}^{\text{atom}}(\mathbf{r}) + V_\text{H}[n_{\text{e}}^{\text{atom}}](\mathbf{r}) +  V_\text{xc}[n^\text{atom}].
\label{eq:AAM Vks}
\end{eqnarray}
How do we calculate each of these terms in practice? The KS potential $V_{\text{KS}}^{\text{atom}}$ is that of the hydrogen atom from a Mermin finite-temperature DFT calculation with a hydrogen pseudopotential. The KS potential includes the electron-nucleus interaction, as well as Hartree and exchange-correlation terms for the atom -- which in the case of hydrogen consist entirely of spurious self-interaction of the single electron. The external potential due to the environment, $V_{\text{ext}}^{\text{env}}$, is the result of convolving the neutral atom KS potential with the ion distribution informed by plasma conditions. This idea is inspired by the neutral pseudoatom molecular dynamics (PAMD) model proposed by Starrett, Daligault, and Saumon \cite{PAMD, Starrett2013}. 

PAMD involves treating a neutral pseudoatom within the average atom model framework and coupling it with classical MD simulations for the ionic structure. This approach allows for the evaluation of structural properties, such as the ion-ion pair correlation function $g_\text{II}(r)$, at a fraction of the cost of DFT-MD methods. The neutral pseudoatom idea \cite{Ziman1967,Perrot1990} is to solve the electron density of a full system $n_\text{e}^\text{full}$ with a nucleus at the origin surrounded by a spherically averaged ionic configuration described by the ion-ion pair correlation function $g_\text{II}(r)$. This same system is then calculated with the central atom removed $n_\text{e}^\text{ext}$; the pseudoatom density is then defined as the difference between these: $n_\text{e}^\text{PA} = n_\text{e}^\text{full} - n_\text{e}^\text{ext}$. This isolates the influence of one nucleus on the electron density. Our electronic model is compatible with the electronic part of the pseudoatom model \textit{sans} the $v^\text{nadd}$ term. Our interest is in investigating the electronic structure effects caused by this term due to strong atomic density overlaps in other compatible ion correlation models such as the PAMD or hypernetted chain approximations for ion closure \cite{Chihara1989,DW_Perrot1982}. The PAMD model is used to generate the $g_\text{II}(r)$ used as an input to our average atom model as below. The ion data was provided by Dr. Charles Starrett from the original work \cite{Starrett2013} and additional calculations run for this work. This $g_\text{II}(r)$ encapsulates the properties of the plasma environment such as the density and temperature. It is used along with the average ion density $n_\text{I}^0$ to get the distribution of ions around a central nucleus, and this is convolved with the atom KS potential as a kernel to produce the environment potential:
\begin{equation}
V_{\text{ext}}^{\text{env}}(r)=\int n_\text{I}(r')V_{\text{KS}}^{\text{atom}}(|\mathbf{r'}-\mathbf{r}|)dr', \qquad
n_\text{I}(r)=n_\text{I}^0g_\text{II}(r)
\end{equation}

Up to this term and the associated Hartree, exchange, and correlation terms, the AAM potential in Eq. \ref{eq:AAM Vtot} is then fully defined and consistent with the pseudoatom model in \cite{Starrett2013} with the distinction that our system is built up from an external $g_\text{II}(r)$ while their pseudoatom model includes this self-consistently. The model can be taken out to arbitrarily large radii which, for practicality, are chosen to be large enough such that the $g_\text{II}(r)$ approaches unity. The total density is that of the isolated atom plus the environment similarly to the pseudoatom density, and the total density is a functional of the atomic density through a similar convolution with the $g_\text{II}(r)$ as for the environment potential. $n_\text{e}^\text{atom}$ and $n_\text{e}^\text{env}$ are the terms required to evaluate the non-additive kinetic potential bi-functional $v^{\text{nadd}}[n_{\text{e}}^{\text{atom}}, n_{\text{e}}^{\text{env}}](r)$, which is the remaining term in the AAM potential: 
\begin{equation}
\begin{aligned}
& n_{\text{e}}^{\text{tot}}=n_\text{e}^\text{atom}+n_\text{e}^\text{env}[n_\text{e}^\text{atom}](r)   \\
& n_\text{e}^\text{env}[n_\text{e}^\text{atom}](r)=\int n_\text{I}(r')n_{\text{e}}^\text{atom}(|\mathbf{r'}-\mathbf{r}|)dr' 
\end{aligned}    
\end{equation}

\subsection{\label{sec:Continuum} Treatment of continuum electrons}

Mermin DFT \cite{Mermin1965} includes finite-temperature effects by occupying electronic states according to the Fermi-Dirac occupation function. Correspondingly, the higher the electronic temperature in Mermin DFT, the more states are needed to appropriately distribute the electrons according to the occupation function at that temperature. This becomes harder with increasing $k_\text{B}T_\text{e}$: as this goes up the system gains more energy, wavefunctions tend to gain more curvature, and to describe this well, a finer spacing is needed. Likewise, as the system ionizes and charge density moves further away from the central nucleus, a larger radius will be needed. Lastly, the number of extra electronic states needed goes up as mentioned; as a general rule of thumb we aim to fill states up to 5$k_\text{B}T_\text{e}$ above the Fermi level and treat the continuum states explicitly:
\begin{equation}
    n_\text{e}(\textbf{r})=\sum_{i=\text{all}} f_i \left| \phi_i(\textbf{r}) \right|^2
\end{equation}
where,
\begin{equation}
    f(\epsilon_i)=\frac{1}{e^{(\epsilon_i-\epsilon_\mathrm{F})/k_{\rm B} T}+1}.
\end{equation}
This becomes quickly intractable and a treatment for the continuum states is needed which does not rely on solving their KS states explicitly, such as the ideal approximation used in \cite{CallowPRR2022}. We will explore this as a direction for future work; no continuum treatment is implemented for this work.

\subsection{\label{sec:parameters} Computational parameters}

We use the LDA exchange-correlation functional \cite{Dirac1930, PBE_LDA1981} with version 0.4 of pseudodojo optimized norm-conserving Vanderbilt pseudopotential \cite{ONCVP2013,pseudodojo2018}. The GGA functional \cite{PBE1996} is expected to give slightly more accurate results and will be considered in future works. For the isolated atom calculations, a grid spacing (in bohrs: $a_0$) of 0.2 $a_0$ is chosen as the converged spacing, and a sphere radius of 30 $a_0$ is used to allow the wavefunctions to go smoothly to zero. For the full average atom calculation,  a grid spacing of 0.5 $a_0$ and a sphere radius of 30 $a_0$ is chosen. The larger spacing was verified to give very close agreement with the 0.2 $a_0$ case in energies, potentials, and electron densities but allows for much faster convergence. The ion temperature $k_\text{B} T_\text{ion}$ is held fixed at $5$ eV and the densities are 0.008, 0.08, 0.8 g/cm$^3$; these properties come into the model from the provided $g_\text{II}(r)$'s. The electron temperature $k_\text{B} T_\text{e}$ is varied from $0$ to $5$ eV to explore the variation across a span of densities and temperatures in $v^\text{nadd}$.

\section{\label{sec:Results}Results}

To obtain results, we follow the above prescription to construct a user-defined potential for Octopus and solve the full average atom with and without an additional $v^{\text{nadd}}$ term as in Eq. \ref{eq:AAM Vtot} -- this allows us to observe its effect in resulting outputs and observables.

\begin{figure}[h!]
    \centering
    \subfloat[\centering ]{ \includegraphics[width=\linewidth]{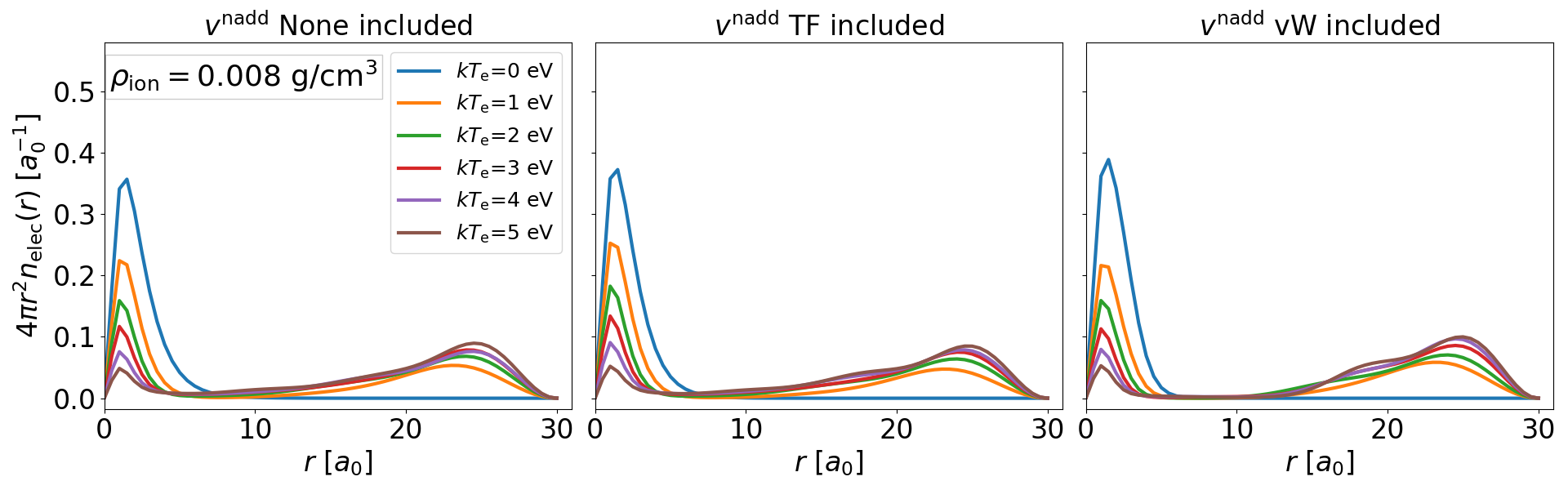} } \\
    \subfloat[\centering ]{ \includegraphics[width=\linewidth]{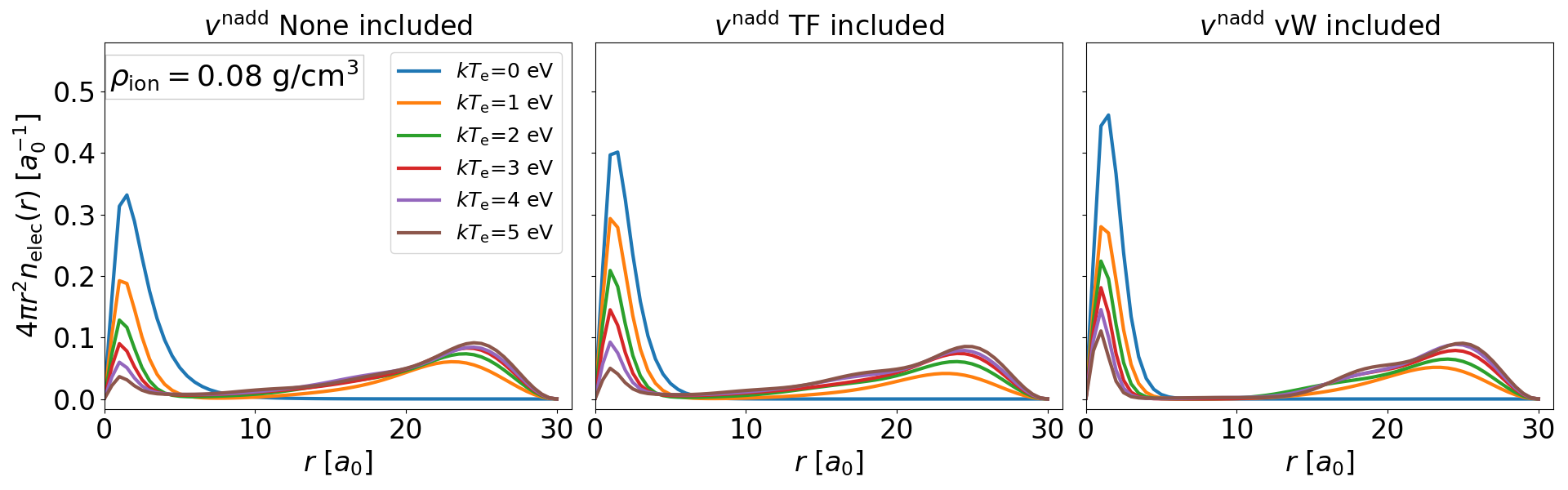} }  \\
    \subfloat[\centering ]{ \includegraphics[width=\linewidth]{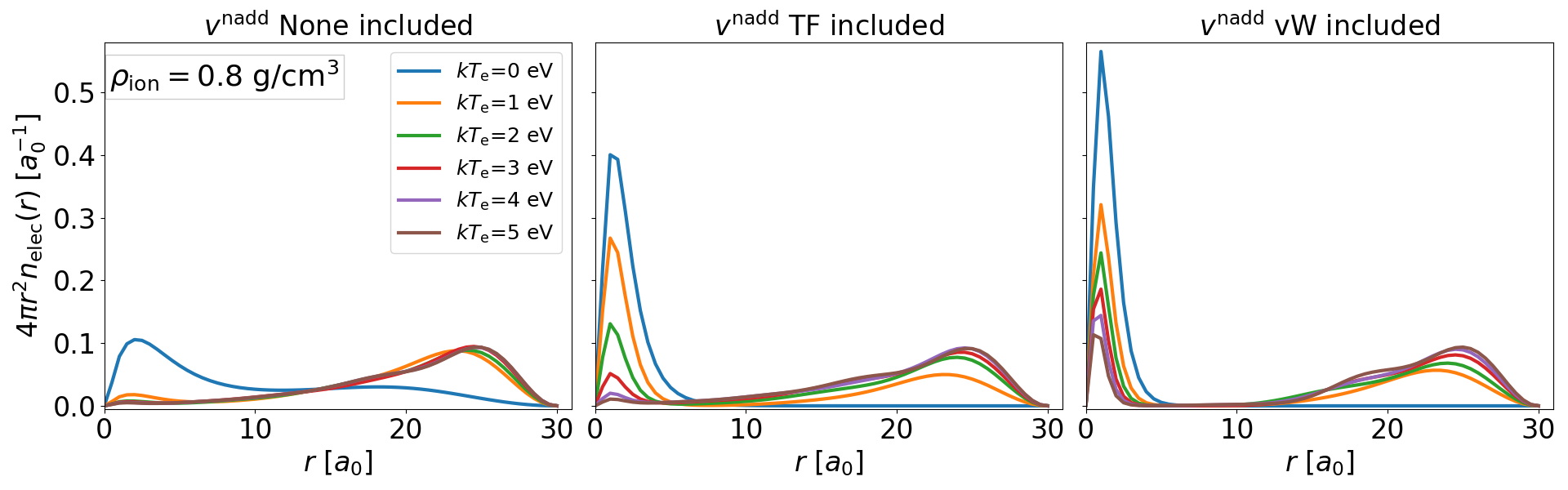} }
    \caption{ Radial distributions $4\pi r^2 n_\text{elec}$ between the system with and without $v^\text{nadd}$, varying electron temperature between 0 to 5 eV, and ion density between 0.008 to 0.8 g/cm$^3$. }
    \label{fig:Vnad_nelec_compare_kT}
\end{figure}

We compare the electron density of the full average atom model and see the effects that arise from including $v^\text{nadd}$ in the AAM. Fig. \ref{fig:Vnad_nelec_compare_kT} shows the electron radial distribution functions around the central atom with varying electron temperature between 0 to 5 eV, ion density between 0.008 to 0.8 g/cm$^3$, and different $v^\text{nadd}$ treatments. Generally, for almost all cases the effect of including $v^\text{nadd}$ is to push the electron density towards the central atom -- this can be interpreted as the effect of the environment pushing in on the central atom due to overlapping densities, which otherwise is not accounted for. Across the temperature and density range, the contribution due to the vW functional is slightly higher than TF. The effect is more prominent for higher densities (bottom row) and is small or negligible for the 0.008 g/cm$^3$ case (top row). For reference, solid density hydrogen is about 0.08 g/cm$^3$.

The trend in temperatures shows that, as the plasma is heated, the $v^\text{nadd}$ effect becomes less prominent. In each plot, as $k_\text{B}T_\text{e}$ increases, the localized density around the central atom reduces, and more density is pushed towards larger radii or unbound states. These plots show the necessity of including a continuum treatment in the future, where the free electrons can be accounted with a much smaller sphere radius which would accelerate the speed of the calculations. Additionally, the clear minima in each radial distribution plot indicate a useful cutoff for the start of the quadratic behavior for continuum states, and thus an approach to calculating the mean ionization state $Z^*$ according to:
\begin{equation}
    Z^*=Z-\int_{r_{\rm min}}^{\infty} 4 \pi r^2 n_\text{elec}(r) \text{d}r,
\end{equation}
where the integral over bound states is approximated by finding the minima in the radial distributions. This approximated $Z^*$ for the various $v^\text{nadd}$ treatments along with the Coulomb coupling parameters are given in Table \ref{tab:plasma_params}. The Coulomb coupling parameter gives a scale of thermal to Coulomb energies and is roughly around one for warm dense matter, and is the regime where our model is most suitable, 
\begin{equation}
    \Gamma=\frac{Z^*{^2}}{k_\text{B}T_\text{ion} R_\text{WS}}.
\end{equation}
\begin{table}[b]
\caption{\label{tab:plasma_params}%
Plasma parameters for each ion density and temperature condition considered
}
\begin{ruledtabular}
\begin{tabular}{llllll}
    $k_\text{B}T_\text{ion}$ [eV] & 
    $\rho$ [g/cm$^3$] & 
    \multicolumn{1}{c}{$\Gamma$}&
    $Z^*$ no $v^\text{nadd}$&
    $Z^*$ TF $v^\text{nadd}$&
    $Z^*$ vW $v^\text{nadd}$\\
    \colrule
    5 & 0.8   & 3.63 & 0.982 & 0.972 & 0.850 \\
    5 & 0.08  & 1.68 & 0.916 & 0.892 & 0.822 \\
    5 & 0.008 & 0.78 & 0.889 & 0.884 & 0.887 \\
\end{tabular}
\end{ruledtabular}
\end{table}

Overall, we observe that $v^\text{nadd}$, though small, has a non-negligible effect on electron density. This effect becomes more prominent with increasing plasma density and less prominent with increasing electron temperature. The increase with plasma density makes sense, as $v^\text{nadd}$ arises from strong density overlaps, and it should approach zero for non-overlapping densities. The reduction of the $v^\text{nadd}$ effect with increasing temperature can be understood by considering the variations in electron density. $v^\text{nadd}$ acts on these variations and has the most significant impact when there are large gradients in $n_\text{elec}$. In contrast, for a fully ionized plasma, where $k_\text{B}T$ far exceeds interaction energies, the plasma density tends to be uniform and flat in all directions. When $v^\text{nadd}$ acts on such a density, it simply produces a constant offset in the total energy, which does not affect observables. As $k_\text{B}T_\text{e}$ increases, the electron density becomes more uniform, the plasma becomes increasingly ionized, and the effect of $v^\text{nadd}$ decreases. We should also expect the contribution of $v^\text{nadd}$ to be greater in systems with more electrons, where orbital overlaps can include core electrons—something that does not occur in a hydrogen atom.

\begin{figure}[H]
    \centering
    \subfloat[\centering ]{ \includegraphics[width=0.33\linewidth]{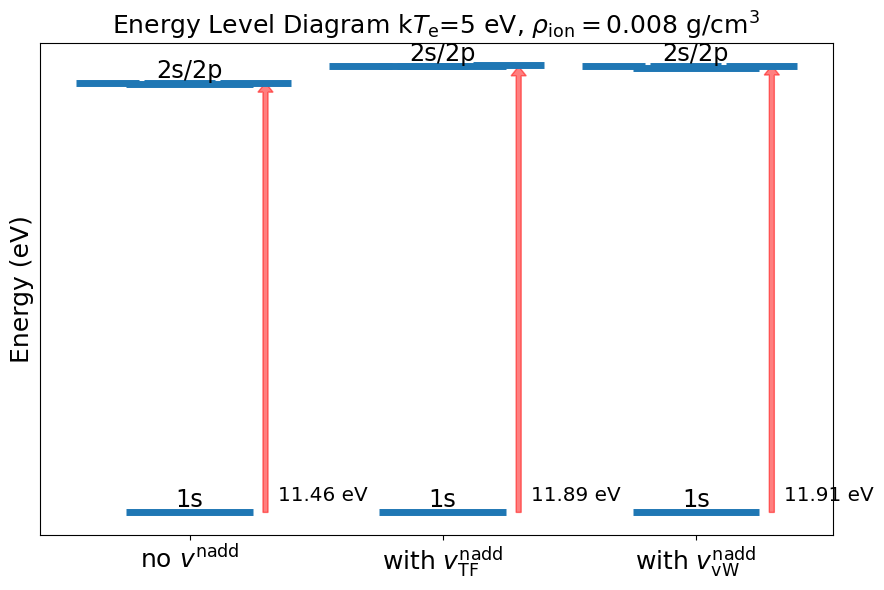} }
    \subfloat[\centering ]{ \includegraphics[width=0.33\linewidth]{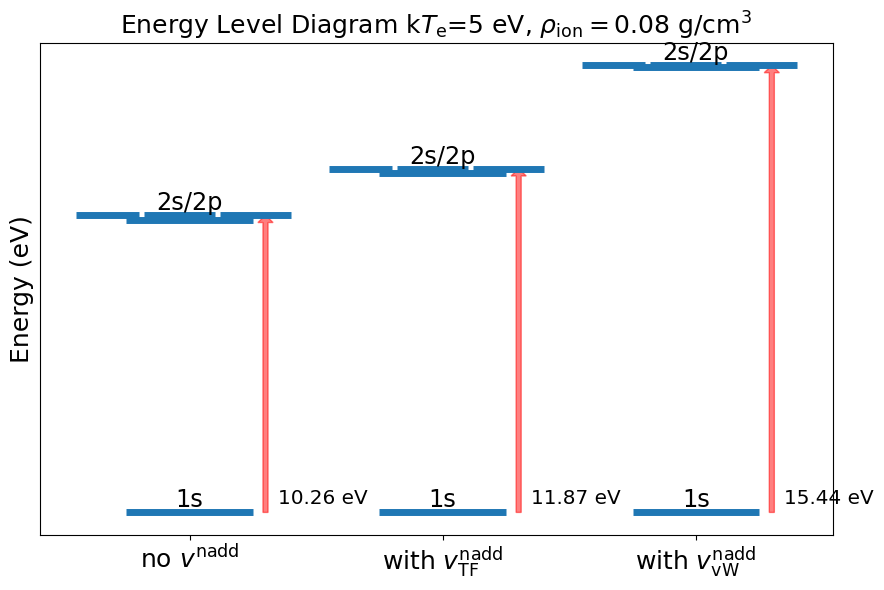} }
    \subfloat[\centering ]{ \includegraphics[width=0.33\linewidth]{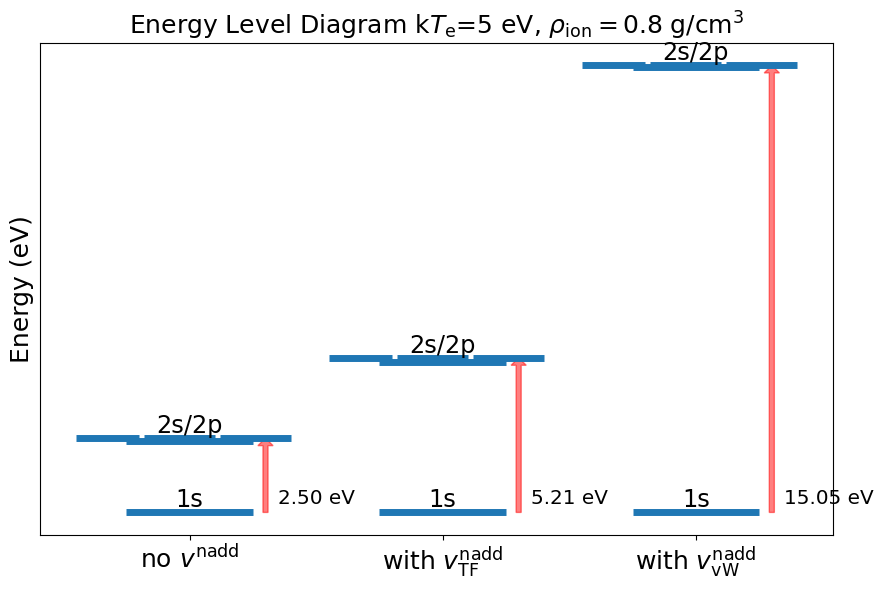} }
    \caption{ Kohn-Sham eigenstates from solving the AAM with and without $v^\text{nadd}$, varying ion density from 0.008 g/cm$^3$ to 0.8 g/cm$^3$, at $k_\text{B}T_\text{ion}=k_\text{B}T_\text{e}=5$ eV. }
    \label{fig:Vnad_Eis}
\end{figure}

While the $v^\text{nadd}$ effects in electron density are interesting and novel to explore, these are very difficult scales to discern experimentally. Instead, we can consider some experimental observables based on the KS eigenvalues. While these eigenvalues of fictitious KS states do not have direct physical meaning, they can serve as crude (under)estimates of the true energy levels whose trends are likely to be correct. Therefore we can look at relative differences between energy levels and how these change with $v^\text{nadd}$. Such shifts can be resolved to sub-eV resolution with X-ray free electron lasers \cite{Dvorak2016}. The energy levels of the first few core states are shown in Fig. \ref{fig:Vnad_Eis}; these are all offset to the 1s level to compare relative differences. The 2s/2p states show a slight splitting which is due to the breaking of radial symmetry in solving the hydrogen atom $1/r$ problem on a Cartesian grid in Octopus; we have not explored any fine structure effects in this study. Instead, we focus on the 1s-2p gap that widens when we include $v^\text{nadd}$ for the higher densities. At the lower density limit, there doesn't seem to be any significant change in the gap which is indicative of very small or zero density overlaps. Again, the effect also appears to be more prominent for the von Weizs\"acker functional compared to Thomas-Fermi; where Thomas-Fermi is more accurate in the limit of high density and many orbitals contributing as in a homogeneous electron gas, and von Weizs\"acker for low density and only one orbital contributing.
 
\begin{figure}[ht]
    \centering
    \includegraphics[width=\linewidth]{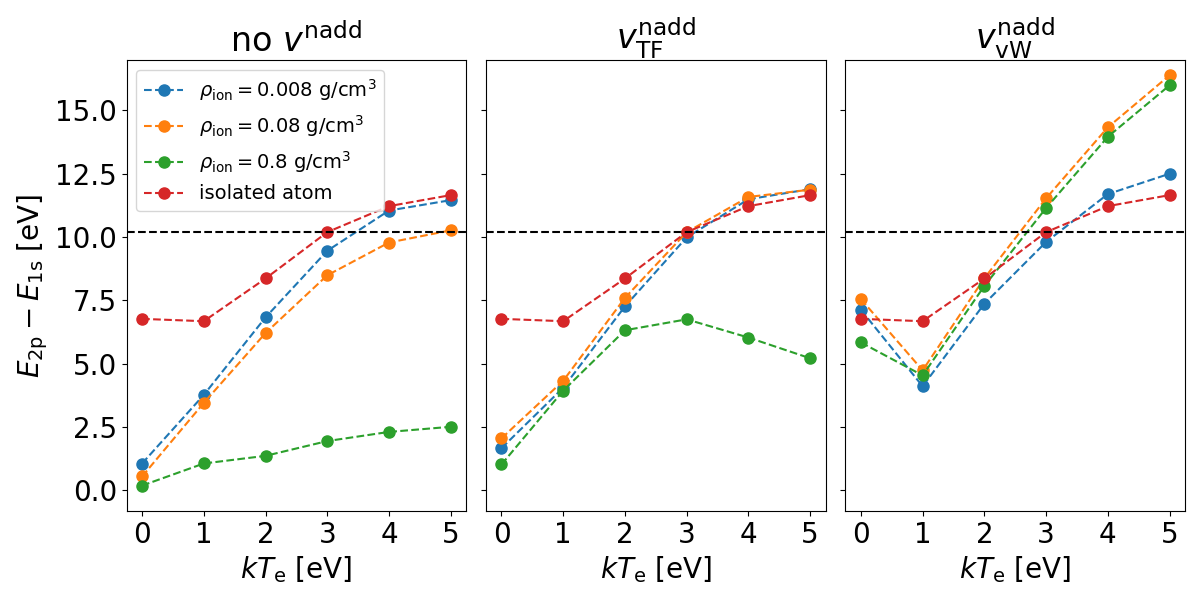}
    \caption{1s-2p gap for the various $v^\text{nadd}$ treatments. ``Isolated atom'' refers to the H atom LDA energies with the same pseudopotential as for the AAM. The black dashed line at $10.2$ eV is the 1s-2p gap known analytically for the H atom, independent of temperature.}
    \label{fig:Vnad_1s-2p_gap}
\end{figure}

The increase in the gap from including $v^\text{nadd}$ can be explained by the energy levels of the simple particle-in-a-box model, $E_n=n^2\pi/2L^2$, which scale inversely with the size of the box $L$. We might be seeing an analogous effect of the system being squeezed ($L$ effectively being reduced) as evident in the electron density plots Fig. \ref{fig:Vnad_nelec_compare_kT} showing the density being pushed towards the central atom. The resultant energy levels are spaced out as the system is confined to a smaller effective $L$. This variation in 1s-2p gap energies with temperature is plotted in Fig. \ref{fig:Vnad_1s-2p_gap} for the different $v^\text{nadd}$ treatments -- the gap increases for increasing $k_\text{B}T_\text{e}$ which indicates increasing repulsion from less localized states of the environment.
In atoms with multiple electronic states, ionization reduces the screening effect on the remaining bound electrons, which leads to an increase in binding energy as ionization increases and as $k_\text{B}T_\text{e}$ rises. While this behavior is nonphysical for a hydrogen atom (as seen in the red curves), it is a real effect for helium and heavier elements. This highlights some of the peculiarities of using DFT for hydrogen. For instance, the LDA gap at $k_\text{B}T_\text{e}=0$ is around 7 eV, whereas the true gap (as known analytically) is closer to 10.2 eV.
Running independent electron calculations in Octopus, which do not include XC effects, yields total energies that agree to within 10 meV and 1s/2p gap energies that agree to sub-meV level with the analytical solutions.

There appear to be a few anomalous points in the trends which again need to be further investigated, for instance with the vW treatment at 1 eV, and the TF 0.8 g/cm$^3$ trend curving up then back down after 3 eV. The leftmost panel shows how the 1s-2p gap evolves for the AAM with no $v^\text{nadd}$ effects -- we can see the gap lowering as the plasma density increases from 0.008 g/cm$^3$ to 0.8 g/cm$^3$ (blue to orange to green). This could be an indicator of continuum lowering which predicts a squeezing together of bound states with increasing pressure \cite{Drake2018,SXHu2017}. When we include the TF or vW treatment of $v^\text{nadd}$, this trend of the gap decreasing with density no longer seems evident. This suggests that the $v^\text{nadd}$ effect of confining the system, which pushes the energy levels apart, reduces continuum lowering which has the effect of squeezing levels together.

\begin{figure}[ht]
    \centering
    \includegraphics[width=
\linewidth]{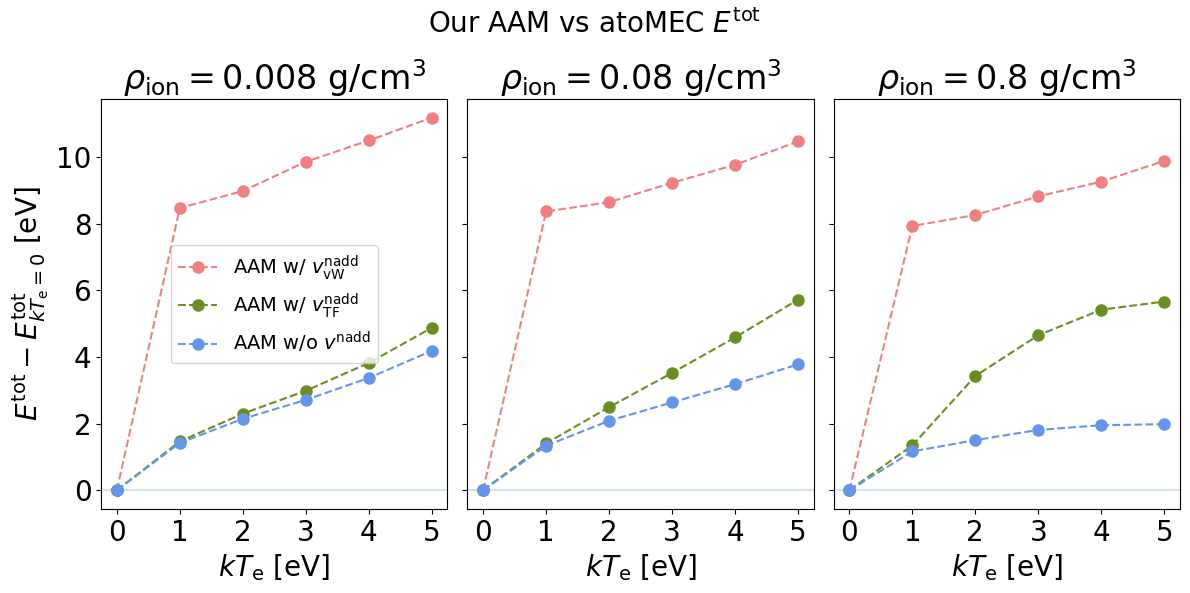}
    \caption{Total energies from our average atom model solved in Octopus with $v^\text{nadd}$.
    Energies are relative to the $k_{\rm B} T_\text{e}=0$ values.
    }
    \label{fig:atoMEC}
\end{figure}

We also compare the variation in total energies with temperature from our model with different $v^\text{nadd}$
(Fig. \ref{fig:atoMEC}). While the total energies are not necessarily measurable experimentally, they are closely related to the pressure, which is. We find $v^\text{nadd}$ can make significant differences in the total energy curve. The temperature trend is consistently increasing in all cases, but as density increases the TF total energy splits off from the no-$v^\text{nadd}$ treatment while the vW variation remains larger than both in all cases. 

\section{\label{sec:Conclusion}Conclusion}

We have identified a missing electronic contribution in existing ion-correlation average atom models which is significant for overlapping electron densities, as occurs for hydrogen in the warm dense matter regime where matter is at typically solid densities and temperatures of several thousand Kelvins. This contribution comes from $v^\text{nadd}$ which accounts for the non-additive part of the potential due to partitioning the exact kinetic energy of the fully interacting system into an embedded and environment contribution. This is applied in a finite-temperature real-space DFT-based AAM and we use approximate kinetic energy functionals to explore the effects due to $v^\text{nadd}$ in our DFT observables. It shows a squeezing of densities towards a central atom embedded in a plasma and a spreading apart of energy levels consistent with this squeezing, and an enhanced variation of total energy vs. temperature. In the limit of small density overlaps, $v^\text{nadd}$ goes to zero and we recover an ion-correlation model with no embedding contribution. In general, average atom models offer significant speed-ups in calculation times while still retaining fair accuracy in evaluating dense plasma properties. However, they struggle to precisely account for electronic interactions due to the averaging over complex environments, and as such there is a need in the field for improved accuracy while maintaining their efficiency. The contribution we have investigated is small but easy to implement since it can be fully defined by the electron density of the average atom. The non-additive kinetic potential from embedding theory can add a small piece of missing physics without compromising the efficiency of the models.

\begin{acknowledgments}
We acknowledge receipt of $g_\text{II}(r)$ data and useful discussions about the PAMD model from Dr. C. E. Starrett. This work was supported with funding by the U.S. Department of Energy, National Nuclear Security Administration, Minority Serving Institution Partnership Program, under Award DE-NA0003984, and the Graduate Student Opportunity Program fellowship at the University of California, Merced. Computing resources were provided by the Pinnacles cluster at the University of California, Merced, supported by National Science Foundation Award OAC-2019144.
\end{acknowledgments}

\bibliography{citations.bib}%

\end{document}